\documentclass[12pt]{article}%
\usepackage{amsfonts}
\usepackage{amssymb}
\usepackage{amsmath}
\usepackage{enumitem}
\usepackage{geometry}
\usepackage{graphicx}
\usepackage{setspace}
\usepackage{natbib}
\RequirePackage[colorlinks,citecolor=blue,urlcolor=blue]{hyperref}
\usepackage{booktabs}%
\usepackage[flushleft]{threeparttable}
\usepackage{tikz}
\usetikzlibrary{arrows,shapes.arrows,shapes.geometric,
	backgrounds,decorations.pathmorphing,positioning,fit,automata,
	shapes.swigs} 
\tikzset{
	>=stealth',
	true/.style={
		rectangle,
		draw=black, very thick,
		text width=6.5em,
		minimum height=2em,
		text centered,
		fill=gray, opacity = 0.5},
	punkt/.style={
		rectangle,
		rounded corners,
		draw=black, very thick,
		text width=6.5em,
		minimum height=2em,
		text centered},
	est/.style={
		circle, minimum size=12mm,
		draw=black, very thick,
		text centered},
	shade/.style={
		circle,
		draw=black, very thick, fill=gray!50,
		text centered},
	weight/.style={
		circle,
		draw=black, very thick,
		text width=6.5em,
		minimum height=2em,
		text centered},
	pil/.style={
		->,
		thick,
		shorten <=2pt,
		shorten >=2pt,},
	double/.style={
		<->,
		thick,
		shorten <=2pt,
		shorten >=2pt,},
	dash/.style={
		dashed,
		thick,
		shorten <=2pt,
		shorten >=2pt,},
	dashdouble/.style={
		<->,
		dashed,
		thick,
		shorten <=2pt,
		shorten >=2pt,}
}

\providecommand{\U}[1]{\protect\rule{.1in}{.1in}}
\newtheorem{theorem}{Theorem}

\newtheorem{lemma}[theorem]{Lemma}

\newtheorem{remark}[theorem]{Remark}

\def\T{ {\mathrm{\scriptscriptstyle T}} }

\begin{document}

\title{ }

\begin{center}
\textit{Original Article}\bigskip

{\LARGE Semiparametric causal mediation analysis with}

{\LARGE unmeasured mediator-outcome confounding}

{\Large Baoluo Sun$^{\ast}$ and Ting Ye$^\dag$} $\ $

$^{\ast}${\large Department of Statistics and Data Science, National University of Singapore\ }

$^{\dag}${\large Department of Biostatistics, University of Washington }

\textbf{Abstract}
\end{center}

\noindent Although the exposure can be randomly assigned in studies of mediation effects, any form of direct intervention on the mediator is often infeasible. As a result, unmeasured mediator-outcome confounding can seldom be ruled out. We propose  semiparametric identification of natural direct and indirect effects  in the presence of unmeasured mediator-outcome confounding by leveraging heteroskedasticity restrictions on the observed data law. For inference, we develop semiparametric estimators that remain consistent under partial misspecifications of the observed data model. We illustrate the proposed estimators  through both simulations and an application to evaluate the effect of self-efficacy on fatigue among health care workers during the COVID-19 outbreak.

KEY WORDS: Causal Inference, multiple robustness, natural direct effect; natural indirect effect; unmeasured confounding.

\bigskip\pagebreak

\section{Introduction}

It is often of interest in the health and social sciences to investigate not only the total effect of a point exposure $A$ on an outcome $Y$, but also the direct and indirect effects operating through a given post-exposure mediating variable $M$. Since the seminal work of \cite{baron1986moderator} in the context of linear structural equation models, the notions of natural direct effect (NDE) and natural indirect effect (NIE) have been formalized in the context of a binary exposure under the potential outcomes framework \citep{robins1992identifiability, 10.5555/2074022.2074073}. The NDE and NIE are particularly useful for understanding the causal mediation mechanism as the sum of these two effects is the average treatment effect of $A$ on $Y$. Under the sequential ignorability assumption  of  no unmeasured confounding for the $A$-$M$, $A$-$Y$ and $M$-$Y$ relationships (See Section \ref{sec:assumptions} for a more formal treatment), the NDE and NIE may be nonparametrically identified from the observed data distribution based on the so-called mediation formula \citep{10.5555/2074022.2074073,vanderweele2009conceptual,imai2010identification,vanderweele2015explanation}. Sequential ignorability is often stated as a conditional version within strata of a set of measured baseline covariates $X$ not affected by the exposure, in the hope that no residual unmeasured confounding remains within strata of the measured covariates. If $X$ is high-dimensional or contains numerous continuous components, then estimation and inference for NDE and NIE typically relies on additional modeling assumptions on the  distribution for the observed data $O=(Y,A,M,X)$ \citep{imai2010identification, tchetgen2012semiparametric,tchetgen2014estimation, vanderweele2015explanation,hines2020robust}. 

A fully parametric approach to evaluating the mediation formula typically entails specifying simple linear models for both $E(Y|M,A,X)$ and $E(M|A,X)$ \citep{vanderweele2009conceptual}, which may be sensitive to model misspecifications. On the other hand, nonparametric inference yields multiply robust estimators which remain consistent and asymptotically normal (CAN) within various strict subsets of the observed data model, which includes the conditional densities $f(M|A,X)$ or $f(A|X)$ \citep{tchetgen2014estimation}. As a compromise between the fully parametric and nonparametric approaches, consider the following semiparametric partially linear outcome and mediator models indexed by $\theta=(\theta_1,\theta_2,\theta_3)^\T \in \rm I\!R^{3}$,
\begin{equation}
\begin{gathered}
\label{eq:structural1}
E(Y|M,A,X;\theta_1,\theta_2,g)=\theta_1 M +\theta_2 A + g(X);\\
E(M|A,X;\theta_3,h)=\theta_3 A +h(X),
\end{gathered}
\end{equation} 
where  the confounding effects of the measured covariates on the outcome and mediator are respectively encoded by the functions $g(\cdot)$ and $h(\cdot)$, which remain unspecified. If $X$ is sufficiently rich so that sequential ignorability holds, then $\theta_2$ and $\theta_1\theta_3$ respectively capture the NDE and NIE per unit change in the exposure \citep{hines2020robust}. Let $\pi(X)\equiv E(A|X)$ denote the treatment propensity score. For estimation of $\theta$, \cite{hines2020robust} considered the $3\times 1$ vector estimating function $\varphi(O;\theta,\pi,g,h)$ with components
\begin{equation}
\begin{aligned}
\label{eq:g}
\varphi_1(O;\theta,\pi,g)&=\{A-\pi(X)\}\{Y-\theta_1 M-\theta_2 A- g(X)\}\\
\varphi_2(O;\theta,h,g)&=\{M-\theta_3 A-h(X)\}\{Y-\theta_1 M-\theta_2 A-g(X)\}\\
\varphi_3(O;\theta,\pi,h)&=\{A-\pi(X)\}\{M-\theta_3 A-h(X)\}.
\end{aligned}
\end{equation} 
An augmented G-estimator \citep{robins1994correcting} of $\theta$ may be constructed as the solution to the empirical moment condition $$n^{-1}\sum^n_{i=1}\varphi(O_i;\theta, \hat{\pi},\hat{g},\hat{h})=0,$$ where $(\hat{\pi},\hat{g},\hat{h})$ is a first-stage estimator of the nuisance parameters under user-specified parametric models. Provided any pair of nuisance parameters in $\{\pi(x),g(x),h(x)\}$ is correctly modeled, \cite{hines2020robust} showed that the resulting augmented G-estimator is CAN for the true value of $\theta$ defined under partially linear model (\ref{eq:structural1}).

\subsection{Motivation and related work}

Although in principle one can rule out unmeasured confounding of the $A$-$M$ and $A$-$Y$ relationships by design when the exposure assignment is randomized (possibly within strata of a known set of measured baseline covariates), it is often infeasible to directly manipulate the mediator. As a result, numerous researchers have developed sensitivity analysis \citep{imai2010general,vanderweele2010bias, tchetgen2012semiparametric, ding2016sharp} and partial identification approaches \citep{sjolander2009bounds, robins2010} to assess the impact of departures from the no unmeasured $M$-$Y$ confounding assumption. Identification of causal mediation mechanisms  under unmeasured $M$-$Y$ confounding can sometimes be achieved through the principal stratification approach \citep{gallop2009mediation, mattei2011augmented} or  by leveraging ancillary variables that satisfy certain exclusion restrictions \citep{ imai2013experimental,burgess2015network,doi:10.1111/rssb.12232}. Another major strand of work in the health sciences uses baseline covariates interacted with random exposure assignment as instrumental variables for the effect of the mediator on the outcome \citep{ten2007causal,doi:10.1002/sim.2891, albert2008mediation,small2012mediation,zheng2015causal}; see also the commentary by \cite{ogburn2012}. 

Recently, there has been a growing interest in  econometrics and health sciences to use higher-order moment restrictions as a source of identification in linear structural models without exclusion restrictions \citep{rigobon2003identification, klein2010estimating, lewbel2012using, ett2020}. To the best of our knowledge, the work of \cite{fulcher2019estimation} is the first to extend this identification framework to causal mediation analysis with unmeasured $M$-$Y$ confounding. They considered identification and estimation of the NIE under structural assumptions which imply the semiparametric partially linear model 
\begin{equation}
\begin{gathered}
\label{eq:structural2}
E(Y|M,A,X,U;\theta_1,\theta_2,g)=\theta_1 M +\theta_2 A + g(X,U);\\
E(M|A,X,U;\theta_3,h)=\theta_3 A +h(X,U),
\end{gathered}
\end{equation} 
where $U$ is a set of unmeasured baseline covariates, not affected by the exposure, that confounds the $M$-$Y$ relationship. The unspecified functions $g(\cdot)$ and $h(\cdot)$ now encode the confounding effects of both measured and unmeasured covariates on the outcome and mediator, respectively. In this paper, we build upon the results of \cite{fulcher2019estimation} and establish identification of $\theta=(\theta_1,\theta_2,\theta_3)^\T$ (and hence both the NDE and NIE) under the partially linear model (\ref{eq:structural2}). Furthermore, similar to \cite{hines2020robust}, we propose augmented G-estimators that remain CAN for the true value of $\theta$ defined by (\ref{eq:structural2}) if any one of three strict subsets of the nuisance parameters lie in user-specified parametric models, including one in which the parametric models for the nuisance parameters considered by \cite{fulcher2019estimation} are correctly specified. This marks a significant improvement in robustness to model misspecification, which is especially useful in observational studies when $X$ contains numerous continuous components.

The rest of the paper is organized as follows. In section \ref{sec:assumptions}, we introduce the formal identification conditions for $\theta$ under partially linear model (\ref{eq:structural2}), and present multiply robust augmented G-estimation 
methods in section \ref{sec:dr}. We evaluate the finite-sample performance of these proposed methods through simulation studies
in section \ref{sec:sim} and illustrate the approach with a real data example in section \ref{sec:app}. We explore several possible extensions including allowing for $A$-$M$ interactions in the outcome model in section \ref{sec:ext} before ending with a brief discussion in section \ref{sec:dis}.

\section{Notation and assumptions}
\label{sec:assumptions}

 We use the potential outcomes framework \citep{neyman1923applications,rubin1974estimating} to define the mediation effects of interest.  Let $M_a$ denote the mediator value that would be observed had the exposure $A$ been set, possibly contrary to fact, to level $a$. Similarly, let $Y_{am}$ denote the potential outcome that would be observed  had $A$ been set to level $a$, and $M$ to $m$. The population NDE and NIE of $A$ on $Y$ comparing two exposure levels $a,a^{\prime}$ are given by $\text{NDE}(a,a^{\prime})\equiv E(Y_{a,M_{a^{\prime}}}-Y_{a^{\prime},M_{a^{\prime}}})$ and $\text{NIE}(a,a^{\prime})\equiv E(Y_{a,M_a}-Y_{a,M_{a^{\prime}}})$ respectively  \citep{vanderweele2015explanation}. The NDE and NIE  are particularly relevant for describing the underlying mechanism by which the exposure operates, as their sum equals the population total effect given by $E(Y_{a,M_a}-Y_{a^{\prime},M_{a^{\prime}}})$. Under the sequential ignorability assumption that  for any $(a,a^{\prime},m)$,
\begin{equation}
\begin{gathered}
\label{eq:si1}
Y_{am}\perp A|X,\quad M_a\perp A|X,\quad Y_{am}\perp M| (A,X),\quad Y_{am}\perp M_{a^{\prime}}|X,
\end{gathered}
\end{equation} 
the mediation effects are nonparametrically identified from the observed data distribution as the functionals
\begin{equation}
\begin{split}
\label{eq:med}
\text{NDE}(a,a^{\prime})&=\iint \{E(Y|a,m,x)-E(Y|a^{\prime},m,x)\}f(m|a^{\prime},x)f(x)dmdx;\\
\text{NIE}(a,a^{\prime})&=\iint E(Y|a,m,x)\{f(m|a,x)-f(m|a^{\prime},x)\}f(x)dmdx,
\end{split}
\end{equation}
for all $a,a^{\prime}$ \citep{10.5555/2074022.2074073,imai2010identification,vanderweele2015explanation}. Evaluation of (\ref{eq:med}) in conjunction with the partially linear model (\ref{eq:structural1}) yields $\text{NDE}(a,a^{\prime})=\theta_2(a-a^{\prime})$ and $\text{NIE}(a,a^{\prime})=\theta_1\theta_3(a-a^{\prime})$; hence the NDE and NIE are identified as long as $\theta$ is identified.

\subsection{Identification of mediation effects under unmeasured $M$-$Y$ confounding}

In practical settings it is often infeasible to randomize or intervene on the mediator. Because any unmeasured $M$-$Y$ confounding that can seldom be ruled out, we will assume that (\ref{eq:si1}) only holds conditional on $(X,U)$, i.e. for any $(a,a^{\prime},m)$,  
\begin{equation}
\begin{gathered}
\label{eq:si2}
Y_{am}\perp A|(X,U),\quad M_a\perp A|(X,U),\quad Y_{am}\perp M| (A,X,U),\quad Y_{am}\perp M_{a^{\prime}}|(X,U).
\end{gathered}
\end{equation}
In addition, we will assume that the exposure is randomly assigned either by design or through some natural experiments, so that 
\begin{equation}
\begin{gathered}
\label{eq:rand}
A\perp U|X,
\end{gathered}
\end{equation}
where $B\perp C|D$ indicates conditional independence of $B$ and $C$ given $D$ \citep{dawid1979conditional}. Figure~\ref{fig1} depicts the causal diagram for such a scenario. Under the latent sequential ignorability assumption (\ref{eq:si2}), it is straightward to verify that the mediation effects are now given by the functionals 
\begin{equation}
\begin{split}
\label{eq:med2}
\text{NDE}(a,a^{\prime})&=\iint \{E(Y|a,m,x,u)-E(Y|a^{\prime},m,x,u)\}f(m|a^{\prime},x,u)f(x,u)dmdxdu;\\
\text{NIE}(a,a^{\prime})&=\iint E(Y|a,m,x,u)\{f(m|a,x,u)-f(m|a^{\prime},x,u)\}f(x,u)dmdxdu.
\end{split}
\end{equation}
Evaluation of (\ref{eq:med2}) in conjunction with the partially linear model (\ref{eq:structural2}) yields $\text{NDE}(a,a^{\prime})=\theta_2(a-a^{\prime})$ and $\text{NIE}(a,a^{\prime})=\theta_1\theta_3(a-a^{\prime})$, a result given in \cite{fulcher2019estimation}. The partially linear model (\ref{eq:structural2}) in conjunction with a randomized exposure that satisfies (\ref{eq:rand}) yield the conditional mean independence restrictions
\begin{equation}
\begin{gathered}
\label{eq:obs}
E\{Y-\theta_1 M -\theta_2 A\rvert A,X\}=E\{g(X,U)|A,X\}=g^{\ast}(X);\\
E\{M-\theta_3 A\rvert A,X\}=E\{h(X,U)|A,X\}= h^{\ast}(X).
\end{gathered}
\end{equation}
The main challenge with identification and estimation based on (\ref{eq:obs}) is that there are two restrictions but three unknown parameters in $\theta$. If $U$ is null and thus $g(X,U)=g^{\ast}(X)$ almost surely, \cite{hines2020robust} derived augmented G-estimators of $\theta$ based on the additional restriction $E\{Y-\theta_1 M -\theta_2 A|A,M,X\}=g^{\ast}(X)$. However, this restriction fails to hold when $U$ is non-null, as $E\{g(X,U)|A,M,X\}$ will remain a function of $(A,M,X)$ due to collider bias at $M$ within strata of $X$ as shown in figure~\ref{fig1}. In this paper we do not impose this restriction, and instead leverage the conditional covariance mean independence restriction
\begin{equation*}
\begin{split}
E[\{M-\theta_3 A-{h}^{\ast}(X)\}\{Y-\theta_1 M\}|A,X]
=E[\text{cov}\{g(X,U),h(X,U)\}|A,X]= \rho(X),
\end{split}
\end{equation*}
which holds under (\ref{eq:structural2}) and (\ref{eq:rand}). We summarize the observed data restrictions below.

\begin{lemma}
\label{lem:ident}
Under partially linear model (\ref{eq:structural2}) and a randomized exposure which satisfies (\ref{eq:rand}), the conditional mean independence restriction
\begin{equation}
\begin{gathered}
\label{eq:res}
E\{\psi(O;\theta,h^{\ast})\rvert A,X\}=E\{\psi(O;\theta,h^{\ast})\rvert X\}
\end{gathered}
\end{equation}
holds almost surely, where $\psi(O;\theta,h^{\ast})$ is a $3\times 1$ vector function with components $\psi_1(O;\theta)=\{Y-\theta_1 M-\theta_2 A\}$, $\psi_2(O;\theta,h^{\ast})=\{M-\theta_3 A-{h}^{\ast}(X)\}\{Y-\theta_1 M\}$ and $\psi_3(O;\theta)=\{M-\theta_3 A\}$.
\end{lemma}  
For identification, we also require (\ref{eq:res}) to have a unique solution for $\theta$. This may be partly justified by the linearity of the first and third components of $\psi(O;\theta,h^{\ast})$. The second component is nonlinear in $\theta$ and therefore require higher moment restrictions for identification. Following \cite{fulcher2019estimation}, we will assume that the observed data distribution satisfies the heteroscedasticity condition that for any pair of exposure values $(a,a^{\prime})$, 
\begin{equation}
\begin{gathered}
\label{eq:het}
\text{var}(M|A=a,X)\neq \text{var}(M|A=a^{\prime},X)\text{ if } a\neq a^{\prime},
\end{gathered}
\end{equation}
almost surely. Condition (\ref{eq:het}) is empirically testable under parametric restrictions \citep{breusch1979simple}, and may be motivated from the linear structural equation \citep{pearl2000models} 
\begin{equation}
\begin{gathered}
\label{eq:lse}
M=\lambda(A,X,U,\epsilon)=\lambda_0(\epsilon)A+\lambda_1(X,U,\epsilon),
\end{gathered}
\end{equation}
where $\lambda_0(\cdot)$, $\lambda_1(\cdot)$ are unspecified functions and $\epsilon$ is a latent error that satisfies $\epsilon \perp (A,X,U)$. We note that (\ref{eq:lse}) implies the mediator partially linear model in (\ref{eq:structural2}). Let $\tilde{\lambda}_0(\epsilon)\equiv \lambda_0(\epsilon)-E\{\lambda_0(\epsilon)\}$ and $\tilde{\lambda}_1(X,U,\epsilon)\equiv \lambda_1(X,U,\epsilon)-E\{\lambda_1(X,U,\epsilon)|X\}$. Then the conditional variance $\text{var}(M|A,X)=E[\{\tilde{\lambda}_0(\epsilon)A+\tilde{\lambda}_1(X,U,\epsilon)\}^2|A,X]$  depends on $A$ provided $\lambda_0(\epsilon)$ depends on the latent source of effect heterogeneity $\epsilon$, which is plausible under a variety of health and social sciences settings \citep{ett2020}. We note that heteroscedasticity does not rule out the null effect $E\{\lambda_0(\epsilon)\}=0$.

\setcounter{theorem}{0}
\begin{theorem}
\label{thm:1}
Under partially linear model (\ref{eq:structural2}) and assumptions (\ref{eq:rand}) and (\ref{eq:het}), the parameter $\theta$ is  identified as the unique solution to $E\{\psi(O;\theta,h^{\ast})\rvert A,X\}=E\{\psi(O;\theta,h^{\ast})\rvert X\}$.
\end{theorem}

\section{Semiparametric inference}
\label{sec:dr}

The conditional mean independence restriction (\ref{eq:res}) implies the following unconditional moment condition for $\theta$,
 \begin{equation}
\begin{split}
\label{eq:plug}
 0=E[\{ A-\pi(X)\}\psi(O;\theta,h^{\ast})],
 \end{split}
\end{equation}
which depends on the unknown nuisance parameters $\pi(x)$ and $h^{\ast}(x)$. The NIE estimator of $\theta_1\theta_3$ proposed by \cite{fulcher2019estimation} may be viewed as solving the empirical versions of only the second and third components in  (\ref{eq:plug}), with both $\pi(x)$ and $h^{\ast}(x)$ estimated parametrically. It is clear that (\ref{eq:plug}) does not have mean zero and therefore fails to identify $\theta$ when either or both parametric models for $\pi(x)$ and $h^{\ast}(x)$ are misspecified. On the other hand, nonparametric estimation of $\pi(x)$ and $h^{\ast}(x)$ is often infeasible due to the curse of dimensionality in moderately sized samples \citep{robins1997toward}, a setting which is of particular relevance when the analyst considers a broad collection of covariates and their functional forms to render condition (\ref{eq:rand}) plausible in observational studies. 

As a remedy, following the augmented G-estimation approach \citep{robins1994correcting}, we propose estimators of $\theta$ which remain CAN if various strict subsets of the nuisance parameters $\eta=\{\pi(x),g^{\ast}(x),{\rho}(x),h^{\ast}(x)\}$ are correctly modeled. To motivate the semiparametric estimators, we derive the influence function of any regular and asymptotically linear estimator of $\theta$ based on (\ref{eq:plug}) when $\{\pi(x),h^{\ast}(x)\}$ is estimated nonparametrically \citep{newey1994asymptotic}. This yields the so-called orthogonal moment condition that is locally robust to the nuisance parameters on which it depends \citep{chernozhukov2020locally}.

 \begin{theorem}
\label{thm:if}
The influence function of any regular and asymptotically linear estimator of $\theta$ based on (\ref{eq:plug}) when $\{\pi(x),h^{\ast}(x)\}$ is estimated nonparametrically is given by $-\Delta^{-1}\tilde{\varphi}(O;\theta,\eta)$, where $\Delta\equiv {E[\{A-\pi(X)\}{\partial}\psi(O;\theta,h^{\ast})/ {\partial \theta}]}$ and $\tilde{\varphi}(O;\theta,\eta)$ is a $3\times 1$ vector function  with components
\begin{equation}
\begin{aligned}
\label{eq:g}
\tilde{\varphi}_1(O;\theta,\pi,g^{\ast})&=\{A-\pi(X)\}\{Y-\theta_1 M-\theta_2 A- g^{\ast}(X)\}\\
\tilde{\varphi}_2(O;\theta,\eta)&=\{A-\pi(X)\}\{M-\theta_3 A-{h}^{\ast}(X)\}\{Y-\theta_1 M-\theta_2 A- g^{\ast}(X)\}-\rho(X)\\
\tilde{\varphi}_3(O;\theta,\pi,h^{\ast})&=\{A-\pi(X)\}\{M-\theta_3 A-h^{\ast}(X)\}.
\end{aligned}
\end{equation} 

 \end{theorem}

\subsection{Multiply robust estimation}

We propose semiparametric estimation of $\theta$ based on the estimating function $\tilde{\varphi}(O;\theta,\eta)$ evaluated under the working parametric models $$\{\pi(x;\eta_{1}),g^{\ast}(x;\eta_2),{\rho}(x;\eta_{3}),h^{\ast}(x;\eta_4):\eta=(\eta^{\T}_{1},\eta^{\T}_{2},\eta^{\T}_{3},\eta^{\T}_{4})^{\T} \in \rm I\!R^{q}\},$$ with $q<\infty$.   Under regularity conditions the proposed semiparametric estimator of $\theta$ is shown to be CAN if one, but not necessarily more than one, of the following model assumptions hold:
\begin{enumerate}
 \item[ $\mathcal{M}_1$:] The models for $\{\pi(x),g^{\ast}(x)\}$ are correct.
 \item[ $\mathcal{M}_2$:] The models for $\{\pi(x),h^{\ast}(x)\}$ are correct.
  \item[ $\mathcal{M}_3$:] The models for $\{g^{\ast}(x),{\rho(x)},h^{\ast}(x)\}$ are correct.
\end{enumerate}
In particular, the model assumption of \cite{fulcher2019estimation} is subsumed under $\mathcal{M}_2$.  For estimation purposes, suppose that $(O_1,...,O_n)$ are independent and identically distributed observations. Let $\hat{E}(\cdot)$ denote the empirical mean operator $\hat{E}\{h(O)\}= n^{-1}\sum_{i=1}^n h(O_i)$. Consider the estimator $\hat{\theta}=(\hat{\theta}_1,\hat{\theta}_2,\hat{\theta}_3)^\T$ which solves
  \begin{equation}
\begin{gathered}
\label{eq:estfun}
 0=\hat{E}\{\tilde{\varphi}(O;\theta,\hat{\eta}(\theta))\},
 \end{gathered}
\end{equation} 
with $\hat{\eta}(\theta)$ solving $0=\hat{E}\{\gamma(O;\theta,\eta)\}$ for a fixed value of $\theta$, where $\gamma(O;\theta,\eta)$ is a $q\times 1$ vector function with the components
\begin{equation}
\begin{aligned}
\label{eq:g}
\gamma_1(O;\eta)&=\{\partial \pi(X;\eta_{1}) /\partial \eta_1\}^\T\{A-\pi(X;\eta_{1})\}\\
\gamma_2(O;\theta,\eta)&=\{\partial g^{\ast}(X;\eta_{2}) /\partial \eta_2\}^\T\{Y-\theta_1 M-\theta_2 A- g^{\ast}(X;\eta_2)\}\\
\gamma_3(O;\theta,\eta)&=\{\partial \rho(X;\eta_{3}) /\partial \eta_3\}^\T[\{M-\theta_3 A-{h}^{\ast}(X;\eta_4)\}\{Y-\theta_1 M-\theta_2 A- g^{\ast}(X;\eta_2)\}-\rho(X;\eta_3)]\\
\gamma_4(O;\theta,\eta)&=\{\partial h^{\ast}(X;\eta_{4}) /\partial \eta_4\}^\T\{M-\theta_3 A-h^{\ast}(X;\eta_4)\}.
\end{aligned}
\end{equation} 

\setcounter{theorem}{1}
\begin{lemma}
\label{lem:nie}
Let $\theta^{\dag}$ denote the unique solution to (\ref{eq:plug}). Then under regularity conditions stated in the Appendix, ${n}^{1/2}(\hat{\theta}-\theta^{\dag})\xrightarrow[]{d}\mathcal{N}(0,\Sigma)$  as $n\to\infty$ in the union model $\{\cup_{j=1}^3 \mathcal{M}_j\}$ (multiple robustness), where $$\Sigma=E\left(\left[E\left\{\frac{\partial}{\partial \theta} \tilde{\Phi}(O;\theta,\bar{\eta}(\theta^{\dag}))\Big\rvert_{\theta=\theta^{\dag}}\right\}^{-1} \tilde{\Phi}(O;\theta^{\dag},\bar{\eta}(\theta^{\dag}))\right]^{\otimes 2}\right),$$
$\bar{\eta}(\theta)$ denotes the probability limit of $\hat{\eta}(\theta)$ and 
$$ \tilde{\Phi}(O;\theta,\eta)= \tilde{\varphi}(O;\theta,\eta)-E\left\{\frac{\partial}{\partial \eta}\tilde{\varphi}(O;\theta,\eta)\right\}E\left\{\frac{\partial}{\partial \eta}\gamma(O;\theta,\eta)\right\}^{-1}\gamma(O;\theta,\eta).$$
\end{lemma}
For inference, a consistent estimator $\hat{\Sigma}$ of $\Sigma$ may be constructed by replacing all expected values with empirical means evaluated at $\hat{\theta}$ and $\hat{\eta}(\hat{\theta})$. Then a 95\% Wald confidence interval for the NDE per unit change in the exposure is found by calculating $\hat{\theta}_2\pm 1.96 \hat{\sigma}_2$, where $\hat{\sigma}_2$ is the square root of the $2^{\text{nd}}$ component of the diagonal of $n^{-1}\hat{\Sigma}$. Similar inference for the NIE per unit change in the exposure may be carried out using the multivariate delta method. Alternatively, nonparametric bootstrap may also be used to obtain estimates of $\Sigma$.

 \begin{remark}
As the nuisance parameters in $\mathcal{M}_k$, $k=1,2,3$ are variation independent, the proposed estimation framework provides the analyst with three genuine opportunities, instead of one, to obtain valid inferences about $\theta$ and functionals thereof, even under partial model misspecifications \citep{robins2001comment}. \cite{10.1111/ectj.12097, chernozhukov2020locally} established general regularity conditions for ${n}^{1/2}$-consistent estimation of finite dimensional parameters of interest based on orthogonal moment functions such as $\tilde{\varphi}(O;\theta,\eta)$,  even when the complexity of the nuisance parameter space for $\eta$ is no longer tractable by standard empirical process
methods (e.g. Vapnik-Chervonenkis and Donsker classes). We plan to pursue this in future work.
 \end{remark}

\section{Simulations}
\label{sec:sim}
We perform simulations to study the pointwise properties of $\hat{\theta}$ and associated confidence intervals. We generate the baseline covariates $X_1\sim \mathcal{N}(0,1)$ and $X_2\sim\mathcal{N}(0,1)$ independently, followed by 
\begin{equation*}
\begin{gathered}
U|X\sim \mathcal{N}\{\mu=1+X_1-0.3X_2, \sigma^2=\exp({-1.2+0.8X_1-0.2X_2})\};\\
A|X\sim \text{Bernoulli}[p=\{1+\exp({1-1.5X_1+0.3X_2})\}^{-1}];\\
M= 1+ (1.5+\epsilon)A +0.5U,\quad Y= 1+A+2M+U;
\end{gathered}
\end{equation*}
where $U$ is an unmeasured factor that induce mediator-outcome dependence, and the latent effect heterogeneity  is generated independently as $\epsilon\sim N(0,1)$. The true NDE and NIE for a unit change of exposure value are $1$ and $3$ respectively. Besides the proposed multiply robust estimator \texttt{MR}, we also implement the propensity score-based estimator \texttt{PS} of \cite{fulcher2019estimation} and the product of coefficients estimator \texttt{BK} of \cite{baron1986moderator} which does not account for unmeasured $M$-$Y$ confounding.  We evaluate the estimators under the following four scenarios to investigate the impact of model misspecification for the nuisance parameters: (i) $\{\pi(x),g^{\ast}(x),{\rho}(x),h^{\ast}(x)\}$ are all correctly modeled; (ii) only $\{\pi(x),g^{\ast}(x)\}$ are correctly modeled; (iii) only $\{\pi(x),h^{\ast}(x)\}$ are correctly modeled and (iv) only $\{g^{\ast}(x),{\rho}(x),h^{\ast}(x)\}$ are correctly modeled. A model is misspecified if the standardized versions of the transformed variables $[\exp({0.5X_1}), 10+X_2/\{1+\exp({X_1})\}]$ are used as regressors  instead of $(X_1,X_2)$. Standard errors are obtained using the empirical sandwich estimator. 

We simulate 1000 replicates with sample size $n=800$ for each scenario and summarize the results for estimation of NDE and NIE for a change of exposure value from $0$ to $1$ in Table~\ref{tab1}. The estimators \texttt{MR} and \texttt{PS} perform similarly to each other in terms of absolute bias and coverage in scenarios (i) and (iii), but \texttt{MR} yields noticeably smaller absolute biases and better coverage than \texttt{PS}  in scenarios (ii) and (iv) where either the model for $\pi(x)$ or $h^{\ast}(x)$ is misspecified. In general \texttt{PS} is less efficient than \texttt{MR}, as the latter incorporates additional regression models that capture the associations between $(Y,M)$ and $(A,X)$. The estimator \texttt{BK} shows large bias with coverage proportions below nominal values across scenarios (i)--(iv), in agreement with theory.

\section{A data application}
\label{sec:app}
We apply the proposed methods in reanalysing an observational study investigating the mediating effect of posttraumatic stress disorder (PTSD) symptoms in the association between self-efficacy and fatigue among health care workers during the COVID-19 outbreak \citep{hou2020self}. The cross-sectional data was collected  between March 13 and 20, 2020, from $n=527$ health care workers in Anqing City, Anhui Province, China, which borders Hubei province, the epicenter of the COVID-19 outbreak.  We refer interested readers to \cite{hou2020self} for further details on the study design. 

For this illustration, the exposure $A$ takes on value 1 if self-efficacy is above the sample median of total scores on the General Self-Efficacy Scale, and $0$ otherwise.  PTSD symptoms ($M$) and fatigue ($Y$) are standardized total scores on the PTSD Checklist-Civilian Version and 14-item Fatigue Scale respectively. The vector of observed baseline covariates $X$ consists of an intercept, age, gender, marital status, education level, work experience (in years) and seniority, as well as level of negative coping dichotomized at the sample median. We specify the working  models $\pi(x;\eta_{1})=\{1+\exp({-\eta^\T_{1} x})\}^{-1}$, $g^{\ast}(x;\eta_{2})=\eta^\T_{2} x$, ${\rho}(x;\eta_{3})=\exp({\eta^\T_{3} x})$ and $h^{\ast}(x;\eta_{4})=\eta^\T_{4} x$ for the nuisance parameters. Due to the limited sample size and because negative coping ($X_{nc}$) has been hypothesised as an important effect modifier of the exposure's effects on both the mediator and outcome  \citep{hou2020self}, we further specify $\theta_1(x;\beta_{1})=\beta^\T_{1}(1,x_{nc})^\T$ and $\theta_3(x;\beta_{3})=\beta^\T_{3}(1,x_{nc})^\T$. The Breusch-Pagan test for heteroskedasticity \citep{breusch1979simple} based on identical working models  for the conditional mediator mean and variance yields a p-value of $8.77\times 10^{-7}$, which indicates that the heteroscedasticity condition (\ref{eq:het}) is plausible. 

Table \ref{tab2} shows various estimates of the NDE and NIE of self-efficacy on fatigue mediated though PTSD symptoms. The regression approach \texttt{BK} of \cite{baron1986moderator} yields a NIE estimate with 95\% confidence interval $-.273\pm .084$. This result suggests a significant mediating effect via PTSD symptoms in reducing fatigue, which is consistent with the original findings by \cite{hou2020self}. The proposed approach \texttt{MR} yields a NIE estimate close to $0$, and the concomitant  95\% confidence interval $.066\pm .308$ includes $0$. This suggests that we cannot rule out a null NIE after accounting for possible unmeasured common causes of PTSD and fatigue. The estimator \texttt{PS} proposed by \cite{fulcher2019estimation} gave similar results.

\section{Extensions}
\label{sec:ext}

\subsection{Exposure-mediator interaction}

In the presence of potential $A$-$M$ interaction in their effects on the outcome, we may consider the following partially linear models  indexed by $\theta=(\theta_1,\theta_2,\zeta,\theta_3)^\T \in \rm I\!R^{4}$, 
\begin{equation}
\begin{gathered}
\label{eq:structural3}
E(Y|M,A,X,U;\theta_1,\theta_2,\zeta,g)=\theta_1 M +\theta_2 A + \zeta AM+ g(X,U);\\
E(M|A,X,U;\theta_3,h)=\theta_3 A +h(X,U),
\end{gathered}
\end{equation} 
where $\zeta$ is a scalar parameter encoding the interaction. If interaction is absent, so that $\zeta=0$, then (\ref{eq:structural3}) reduces to (\ref{eq:structural2}). Evaluation of (\ref{eq:med}) in conjunction with (\ref{eq:structural3}) yields $\text{NDE}(a,a^{\prime})=[\theta_2+\zeta\{\theta_3 a^{\prime}+ E(h^{\ast}(X))\}](a-a^{\prime})$ and $\text{NIE}(a,a^{\prime})=\theta_3(\theta_1+\zeta a)(a-a^{\prime})$ \citep{vanderweele2015explanation}. Because the moment condition
\begin{equation}
\begin{split}
\label{eq:plug2}
 0=E[d(X)\{ A-\pi(X)\}\psi(O;\theta,h^{\ast})],
 \end{split}
\end{equation}
holds for an arbitrary $4\times 1$ vector function $d(x)$, $\theta$ is identified via (\ref{eq:plug2}) if $${E[d(X)\{A-\pi(X)\}{\partial}\psi(O;\theta,h^{\ast})/ {\partial \theta}]}$$ is non-singular. A multiply robust estimator of $\theta$ may then be constructed based on the empirical version of the augmented G-estimation moment condition $0=E[d(X)\tilde{\varphi}(O;\theta,\eta)]$. However, because of its dependence on $h^{\ast}(x)$, the proposed estimator of $\text{NDE}(a,a^{\prime})$ can only be doubly robust in the union model $\{\mathcal{M}_2\cup\mathcal{M}_3\}$.

\subsection{Binary mediator}

 The proposed semiparametric framework also extends to a binary mediator under the following log-linear model, 
 
   \begin{equation}
\begin{split}
\label{eq:loglinear}
E(Y|M,A,X,U;\theta_1,\theta_2,g)&=\theta_1 M+\theta_2 A+g(X,U);\\
\log\{p( M=1|A,X,U;\theta_3,h)\}&=\theta_3A +h(X,U).
\end{split}
\end{equation}
Evaluation of (\ref{eq:med2}) in conjunction with the log-linear model (\ref{eq:loglinear}) yields $\text{NDE}(a,a^{\prime})=\theta_2(a-a^{\prime})$ and $\text{NIE}(a,a^{\prime})=\theta_1\{\exp({\theta_3a})-\exp({\theta_3a^{\prime}})\}E\{\tilde{h}(X)\}$, where $$\tilde{h}(X)= E[Me^{-\theta_3 A}|X]=E[\exp\{h(X,U)\}|X].$$ Under the log-linear model (\ref{eq:loglinear}) and a randomized exposure which satisfies (\ref{eq:rand}), the conditional mean independence restriction \begin{equation}
\begin{gathered}
\label{eq:betarestriction2}
E\{\tilde{\psi}(O;\theta,\tilde{h}) \rvert A,X\}=E\{\tilde{\psi}(O;\theta,\tilde{h}) \rvert X\},
\end{gathered}
\end{equation}
 holds almost surely, where $\tilde{\psi}(O;\theta,\tilde{h})$ is a $3\times 1$ vector function with components $\tilde{\psi}_1(O;\theta)=\{Y-\theta_1 M-\theta_2 A\}$, $\tilde{\psi}_2(O;\theta,\tilde{h})=\{Y-\theta_1 M\}\{M\exp({-\theta_3 A})-\tilde{h}(X)\}$ and $\tilde{\psi}_3(O;\theta)=\{M\exp({-\theta_3 A})\}$. Augmented G-estimation of $\theta$ may proceed based on identifying restriction (\ref{eq:betarestriction2}), using methods analogous to those described in section \ref{sec:dr}. For binary outcome, a direct extension of existing methods  to identify NDE and NIE on the risk ratio scale generally require the conditional density $f(M|A,X,U)$ to be normal with constant variance \citep{vanderweele2010odds,valeri2013mediation, vanderweele2015explanation}, in which case the heteroscedasticity condition (\ref{eq:het})  fails to hold. A possible direction is to identify and estimate NDE and NIE on the risk difference scale under the proposed framework. 

\section{Discussions}
\label{sec:dis}

Unmeasured $M$-$Y$ confounding is particularly pernicious for credible causal mediation analysis in the health and social sciences, as the mediator can seldom be directly manipulated. The main contribution of this paper is a robust inference framework for the NDE and NIE  under unmeasured $M$-$Y$ confounding in partially linear models by leveraging heteroscedasticity of $M$ with respect to $A$, a condition which is empirically testable. It should be pointed out here that the fourth condition $Y_{am}\perp M_{a^{\ast}}|(X,U)$ in (\ref{eq:si2}) cannot be guaranteed through experimental interventions, even if we were able to randomize both the exposure and the mediator \citep{10.5555/3020419.3020437, imai2013experimental, robins2010alternative}. In the absence of exposure-induced confounding, dropping the fourth condition from (\ref{eq:si2}) imbues the functionals in (\ref{eq:med2}) with alternative causal interpretations as interventional analogues of the NDE and NIE \citep{vanderweele2014effect, vanderweele2015explanation}. Because all our results in this paper concern identification and estimation of the functionals in (\ref{eq:med2}),  they can be readily applied under this alternative interpretation. The proposed framework  can also be extended in several other important directions, including mediation analysis with survival data and multiple mediators under unmeasured mediator-outcome confounding, which we plan to pursue in future research.

\section*{Acknowledgments}

BaoLuo Sun is supported by the National University of Singapore Start-Up Grant (R-155-000-203-133). The authors thank Drs. Eric Tchetgen Tchetgen and Xu Shi for their insightful comments.

\bibliographystyle{apalike}
\bibliography{paper-ref}

\begin{thebibliography}{}

\bibitem[Albert, 2008]{albert2008mediation}
Albert, J.~M. (2008).
\newblock Mediation analysis via potential outcomes models.
\newblock {\em Statistics in medicine}, 27(8):1282--1304.

\bibitem[Baron and Kenny, 1986]{baron1986moderator}
Baron, R.~M. and Kenny, D.~A. (1986).
\newblock The moderator--mediator variable distinction in social psychological
  research: Conceptual, strategic, and statistical considerations.
\newblock {\em Journal of personality and social psychology}, 51(6):1173.

\bibitem[Breusch and Pagan, 1979]{breusch1979simple}
Breusch, T.~S. and Pagan, A.~R. (1979).
\newblock A simple test for heteroscedasticity and random coefficient
  variation.
\newblock {\em Econometrica: Journal of the Econometric Society}, pages
  1287--1294.

\bibitem[Burgess et~al., 2015]{burgess2015network}
Burgess, S., Daniel, R.~M., Butterworth, A.~S., Thompson, S.~G., and
  Consortium, E.-I. (2015).
\newblock Network mendelian randomization: using genetic variants as
  instrumental variables to investigate mediation in causal pathways.
\newblock {\em International journal of epidemiology}, 44(2):484--495.

\bibitem[Chernozhukov et~al., 2018]{10.1111/ectj.12097}
Chernozhukov, V., Chetverikov, D., Demirer, M., Duflo, E., Hansen, C., Newey,
  W., and Robins, J. (2018).
\newblock {Double/debiased machine learning for treatment and structural
  parameters}.
\newblock {\em The Econometrics Journal}, 21(1):C1--C68.

\bibitem[Chernozhukov et~al., 2020]{chernozhukov2020locally}
Chernozhukov, V., Escanciano, J.~C., Ichimura, H., Newey, W.~K., and Robins,
  J.~M. (2020).
\newblock Locally robust semiparametric estimation.

\bibitem[Dawid, 1979]{dawid1979conditional}
Dawid, A.~P. (1979).
\newblock Conditional independence in statistical theory.
\newblock {\em Journal of the Royal Statistical Society: Series B
  (Methodological)}, 41(1):1--15.

\bibitem[Didelez et~al., 2006]{10.5555/3020419.3020437}
Didelez, V., Dawid, A.~P., and Geneletti, S. (2006).
\newblock Direct and indirect effects of sequential treatments.
\newblock In {\em Proceedings of the Twenty-Second Conference on Uncertainty in
  Artificial Intelligence}, UAI'06, pages 138--146, Arlington, Virginia, USA.
  AUAI Press.

\bibitem[Ding and Vanderweele, 2016]{ding2016sharp}
Ding, P. and Vanderweele, T.~J. (2016).
\newblock Sharp sensitivity bounds for mediation under unmeasured
  mediator-outcome confounding.
\newblock {\em Biometrika}, 103(2):483--490.

\bibitem[Dunn and Bentall, 2007]{doi:10.1002/sim.2891}
Dunn, G. and Bentall, R. (2007).
\newblock Modelling treatment-effect heterogeneity in randomized controlled
  trials of complex interventions (psychological treatments).
\newblock {\em Statistics in Medicine}, 26(26):4719--4745.

\bibitem[Fr\"{o}lich and Huber, 2017]{doi:10.1111/rssb.12232}
Fr\"{o}lich, M. and Huber, M. (2017).
\newblock Direct and indirect treatment effects-causal chains and mediation
  analysis with instrumental variables.
\newblock {\em Journal of the Royal Statistical Society: Series B (Statistical
  Methodology)}, 79(5):1645--1666.

\bibitem[Fulcher et~al., 2019]{fulcher2019estimation}
Fulcher, I.~R., Shi, X., and Tchetgen~Tchetgen, E.~J. (2019).
\newblock Estimation of natural indirect effects robust to unmeasured
  confounding and mediator measurement error.
\newblock {\em Epidemiology}, 30(6):825--834.

\bibitem[Gallop et~al., 2009]{gallop2009mediation}
Gallop, R., Small, D.~S., Lin, J.~Y., Elliott, M.~R., Joffe, M., and Ten~Have,
  T.~R. (2009).
\newblock Mediation analysis with principal stratification.
\newblock {\em Statistics in medicine}, 28(7):1108--1130.

\bibitem[Hines et~al., 2021]{hines2020robust}
Hines, O., Vansteelandt, S., and Diaz-Ordaz, K. (2021).
\newblock Robust inference for mediated effects in partially linear models.
\newblock {\em Psychometrika}, pages 1--24.

\bibitem[Hou et~al., 2020]{hou2020self}
Hou, T., Dong, W., Zhang, R., Song, X., Zhang, F., Cai, W., Liu, Y., and Deng,
  G. (2020).
\newblock Self-efficacy and fatigue among health care workers during covid-19
  outbreak: A moderated mediation model of posttraumatic stress disorder
  symptoms and negative coping.
\newblock {\em Preprint}.

\bibitem[Imai et~al., 2010a]{imai2010general}
Imai, K., Keele, L., and Tingley, D. (2010a).
\newblock A general approach to causal mediation analysis.
\newblock {\em Psychological methods}, 15(4):309.

\bibitem[Imai et~al., 2010b]{imai2010identification}
Imai, K., Keele, L., and Yamamoto, T. (2010b).
\newblock Identification, inference and sensitivity analysis for causal
  mediation effects.
\newblock {\em Statistical science}, pages 51--71.

\bibitem[Imai et~al., 2013]{imai2013experimental}
Imai, K., Tingley, D., and Yamamoto, T. (2013).
\newblock Experimental designs for identifying causal mechanisms.
\newblock {\em Journal of the Royal Statistical Society: Series A (Statistics
  in Society)}, 176(1):5--51.

\bibitem[Klein and Vella, 2010]{klein2010estimating}
Klein, R. and Vella, F. (2010).
\newblock Estimating a class of triangular simultaneous equations models
  without exclusion restrictions.
\newblock {\em Journal of Econometrics}, 154(2):154--164.

\bibitem[Lewbel, 2012]{lewbel2012using}
Lewbel, A. (2012).
\newblock Using heteroscedasticity to identify and estimate mismeasured and
  endogenous regressor models.
\newblock {\em Journal of Business \& Economic Statistics}, 30(1):67--80.

\bibitem[Mattei and Mealli, 2011]{mattei2011augmented}
Mattei, A. and Mealli, F. (2011).
\newblock Augmented designs to assess principal strata direct effects.
\newblock {\em Journal of the Royal Statistical Society: Series B (Statistical
  Methodology)}, 73(5):729--752.

\bibitem[Newey, 1994]{newey1994asymptotic}
Newey, W.~K. (1994).
\newblock The asymptotic variance of semiparametric estimators.
\newblock {\em Econometrica: Journal of the Econometric Society}, pages
  1349--1382.

\bibitem[Neyman, 1923]{neyman1923applications}
Neyman, J. (1923).
\newblock Sur les applications de la th{\'e}orie des probabilit{\'e}s aux
  experiences agricoles: Essai des principes.
\newblock {\em Roczniki Nauk Rolniczych}, 10:1--51.

\bibitem[Ogburn, 2012]{ogburn2012}
Ogburn, E.~L. (2012).
\newblock Commentary of ``mediation analysis without sequential ignorability:
  using baseline covariates interacted with random assignment as instrumental
  variables''.
\newblock {\em Journal of Statistical Research}, 46:105--111.

\bibitem[Pearl, 2000]{pearl2000models}
Pearl, J. (2000).
\newblock {\em Causality: Models, Reasoning and Inference}.
\newblock Cambridge, UK: Cambridge University Press.

\bibitem[Pearl, 2001]{10.5555/2074022.2074073}
Pearl, J. (2001).
\newblock Direct and indirect effects.
\newblock In {\em Proceedings of the Seventeenth Conference on Uncertainty in
  Artificial Intelligence}, UAI'01, pages 411--420, San Francisco, CA, USA.
  Morgan Kaufmann Publishers Inc.

\bibitem[Rigobon, 2003]{rigobon2003identification}
Rigobon, R. (2003).
\newblock Identification through heteroskedasticity.
\newblock {\em Review of Economics and Statistics}, 85(4):777--792.

\bibitem[Robins, 1994]{robins1994correcting}
Robins, J.~M. (1994).
\newblock Correcting for non-compliance in randomized trials using structural
  nested mean models.
\newblock {\em Communications in Statistics-Theory and methods},
  23(8):2379--2412.

\bibitem[Robins and Greenland, 1992]{robins1992identifiability}
Robins, J.~M. and Greenland, S. (1992).
\newblock Identifiability and exchangeability for direct and indirect effects.
\newblock {\em Epidemiology}, pages 143--155.

\bibitem[Robins et~al., 1992]{robins1992estimating}
Robins, J.~M., Mark, S.~D., and Newey, W.~K. (1992).
\newblock Estimating exposure effects by modelling the expectation of exposure
  conditional on confounders.
\newblock {\em Biometrics}, pages 479--495.

\bibitem[Robins and Richardson, 2010a]{robins2010}
Robins, J.~M. and Richardson, T. (2010a).
\newblock Alternative graphical causal models and the identification of direct
  effects.
\newblock In Shrout, P., Keyes, K., and Ornstein, K., editors, {\em Causality
  and Psychopathology: Finding the Determinants of Disorders and Their Cures},
  pages 103--58. Oxford University Press, Oxford, UK.

\bibitem[Robins and Richardson, 2010b]{robins2010alternative}
Robins, J.~M. and Richardson, T.~S. (2010b).
\newblock Alternative graphical causal models and the identification of direct
  effects.
\newblock {\em Causality and psychopathology: Finding the determinants of
  disorders and their cures}, pages 103--158.

\bibitem[Robins and Ritov, 1997]{robins1997toward}
Robins, J.~M. and Ritov, Y. (1997).
\newblock Toward a curse of dimensionality appropriate (coda) asymptotic theory
  for semi-parametric models.
\newblock {\em Statistics in medicine}, 16(3):285--319.

\bibitem[Robins and Rotnitzky, 2001]{robins2001comment}
Robins, J.~M. and Rotnitzky, A. (2001).
\newblock Comment on the bickel and kwon article,``inference for semiparametric
  models: Some questions and an answer''.
\newblock {\em Statistica Sinica}, 11(4):920--936.

\bibitem[Rubin, 1974]{rubin1974estimating}
Rubin, D.~B. (1974).
\newblock Estimating causal effects of treatments in randomized and
  nonrandomized studies.
\newblock {\em Journal of educational Psychology}, 66(5):688.

\bibitem[Sj{\"o}lander, 2009]{sjolander2009bounds}
Sj{\"o}lander, A. (2009).
\newblock Bounds on natural direct effects in the presence of confounded
  intermediate variables.
\newblock {\em Statistics in Medicine}, 28(4):558--571.

\bibitem[Small, 2012]{small2012mediation}
Small, D.~S. (2012).
\newblock Mediation analysis without sequential ignorability: using baseline
  covariates interacted with random assignment as instrumental variables.
\newblock {\em Journal of Statistical Research}, 46:91--103.

\bibitem[Tchetgen~Tchetgen and Shpitser, 2012]{tchetgen2012semiparametric}
Tchetgen~Tchetgen, E.~J. and Shpitser, I. (2012).
\newblock Semiparametric theory for causal mediation analysis: efficiency
  bounds, multiple robustness, and sensitivity analysis.
\newblock {\em Annals of statistics}, 40(3):1816.

\bibitem[Tchetgen~Tchetgen and Shpitser, 2014]{tchetgen2014estimation}
Tchetgen~Tchetgen, E.~J. and Shpitser, I. (2014).
\newblock Estimation of a semiparametric natural direct effect model
  incorporating baseline covariates.
\newblock {\em Biometrika}, 101(4):849--864.

\bibitem[Tchetgen~Tchetgen et~al., 2020]{ett2020}
Tchetgen~Tchetgen, E.~J., Sun, B., and Walter, S. (2020).
\newblock The genius approach to robust mendelian randomization inference.
\newblock {\em Statistical Science}, in press.

\bibitem[Ten~Have et~al., 2007]{ten2007causal}
Ten~Have, T.~R., Joffe, M.~M., Lynch, K.~G., Brown, G.~K., Maisto, S.~A., and
  Beck, A.~T. (2007).
\newblock Causal mediation analyses with rank preserving models.
\newblock {\em Biometrics}, 63(3):926--934.

\bibitem[Valeri and VanderWeele, 2013]{valeri2013mediation}
Valeri, L. and VanderWeele, T.~J. (2013).
\newblock Mediation analysis allowing for exposure--mediator interactions and
  causal interpretation: theoretical assumptions and implementation with sas
  and spss macros.
\newblock {\em Psychological methods}, 18(2):137.

\bibitem[VanderWeele, 2010]{vanderweele2010bias}
VanderWeele, T.~J. (2010).
\newblock Bias formulas for sensitivity analysis for direct and indirect
  effects.
\newblock {\em Epidemiology (Cambridge, Mass.)}, 21(4):540.

\bibitem[VanderWeele, 2015]{vanderweele2015explanation}
VanderWeele, T.~J. (2015).
\newblock {\em Explanation in causal inference: methods for mediation and
  interaction}.
\newblock Oxford University Press.

\bibitem[VanderWeele and Vansteelandt, 2009]{vanderweele2009conceptual}
VanderWeele, T.~J. and Vansteelandt, S. (2009).
\newblock Conceptual issues concerning mediation, interventions and
  composition.
\newblock {\em Statistics and its Interface}, 2(4):457--468.

\bibitem[VanderWeele and Vansteelandt, 2010]{vanderweele2010odds}
VanderWeele, T.~J. and Vansteelandt, S. (2010).
\newblock Odds ratios for mediation analysis for a dichotomous outcome.
\newblock {\em American journal of epidemiology}, 172(12):1339--1348.

\bibitem[VanderWeele et~al., 2014]{vanderweele2014effect}
VanderWeele, T.~J., Vansteelandt, S., and Robins, J.~M. (2014).
\newblock Effect decomposition in the presence of an exposure-induced
  mediator-outcome confounder.
\newblock {\em Epidemiology (Cambridge, Mass.)}, 25(2):300.

\bibitem[Ye et~al., 2021]{ye2021geniusmawii}
Ye, T., Liu, Z., Sun, B., and Tchetgen, E.~T. (2021).
\newblock Genius-mawii: For robust mendelian randomization with many weak
  invalid instruments.

\bibitem[Zheng and Zhou, 2015]{zheng2015causal}
Zheng, C. and Zhou, X.-H. (2015).
\newblock Causal mediation analysis in the multilevel intervention and
  multicomponent mediator case.
\newblock {\em Journal of the Royal Statistical Society: Series B: Statistical
  Methodology}, pages 581--615.

\end{thebibliography}

\newpage

\section*{Appendix }
\label{app:theorem}
\renewcommand{\theequation}{A\arabic{equation}}
\setcounter{equation}{0} 

\subsection*{Proof of Lemma 2.1}
We make repeated use of the following equalities in this and subsequent proofs. For any arbitrary functions $\sigma_1(X)$ and $\sigma_2(A,X)$ and at the true value for $\theta$,
\begin{eqnarray}
	\label{identity_1}   && E[\sigma_1(X)\{A-\pi(X)\}]=0;\\
	\label{identity_3}   && E[\sigma_2(A,X)\{Y-\theta_1 M-\theta_2 A-g^{\ast}(X)\}]  = 0;\\
	\label{identity_2}   && E[\sigma_2(A,X)\{M-\theta_3 A-h^{\ast}(X)\}]  = 0.
\end{eqnarray}
We have
 \begin{equation*}
\begin{split}
E&[\{Y-\theta_1 M\}\{M-\theta_3 A-h^{\ast}(X)\}|A,X]\\
=&E[\{Y-\theta_1 M-\theta_2 A-g^{\ast}(X)\}\{M-\theta_3 A-h^{\ast}(X)\}|A,X] \quad (\text{\ref{identity_2}})\\
=&E[\{g(X,U)-g^{\ast}(X)\}\{h(X,U)-h^{\ast}(X)\}|A,X]\\
=&\text{cov}[g(X,U),h(X,U)|A,X]=\text{cov}[g(X,U),h(X,U)|X],
\end{split}
 \end{equation*}
where the second equality follows from taking iterated expectations under partially linear model (1.3) with respect to $(A,M,X,U)$ followed by $(A,X,U)$, and the last equality  holds by (2.7).

\subsection*{Proof of Theorem 2.1}
{
Let $m(A,X;\theta,h^{\ast})\equiv E\{\psi(O;\theta,h^{\ast})\rvert A,X\}-E\{\psi(O;\theta,h^{\ast})\rvert X\}$. Lemma 2.1 implies that  $m(A,X;\theta,h^{\ast})=0$ almost surely at the truth $\theta$.  Suppose another parameter value $\theta^{\star}\in{\rm I\!R}^3$ satisfies $m(A,X;\theta^{\star},h^{\ast})=0$ almost surely.  Then $m(A,X;\theta,h^{\ast})-m(A,X;\theta^{\star},h^{\ast})=\tilde{m}(A,X;\theta^{\star}-\theta,h^{\ast})=0$ almost surely, where the $3\times 1$ vector function $\tilde{m}(A,X;\theta^{\star}-\theta,h^{\ast})$ has the components
 \begin{equation*}
\begin{split}
\tilde{m}_1(A,X;\theta^{\star}-\theta,h^{\ast})&=(\theta^{\star}_1-\theta_1)r_1(A,X)+(\theta^{\star}_2-\theta_2)\{A-\pi(X)\}\\
\tilde{m}_2(A,X;\theta^{\star}-\theta,h^{\ast})&=(\theta^{\star}_1-\theta_1)[\text{var}(M|A,X)-E\{\text{var}(M|A,X)|X\}]+(\theta^{\star}_3-\theta_3)r_2(A,X)\\
\tilde{m}_3(A,X;\theta^{\star}-\theta,h^{\ast})&=(\theta^{\star}_3-\theta_3)\{A-\pi(X)\}.
\end{split}
 \end{equation*}
The heteroscedasticity condition (2.11) implies $\theta^{\star}-\theta=0$, provided $f(A|X)$ is non-degenerate.

\subsection*{Proof of Theorem 3.1}

Let $\{F_{t}(O)\in \mathcal{M}\}$ denote a path through the observed data distribution, with true distribution $F_0$ at $t=0$ and the corresponding score $\partial \ln(dF_{t})/\partial {t}|_{{t}=0}=S(O)=S(Y|M,A,X)+S(M|A,X)+S(A|X)+S(X)$. We calculate the pathwise derivative $\partial \theta_{t}/\partial {t}|_{{t}=0}$ based on the moment condition $0=E_{t} [\{A-\pi_t(X)\} \psi(O; \theta_{t},h^{\ast}_{t})],$ where $E_{t}[\cdot]$ denotes the expectation at $F_{t}$. Differentiating under the integral yields $$\partial \theta_{t}/\partial {t}\vert_{t=0}=-\Delta^{-1}[E\{\psi(O;\theta,h^{\ast})S(O)\}+J_{\pi}+J_h],$$  where $J_{\pi}\equiv\partial E [ \{A-\pi_t(X)\}\psi(O; \theta,h^{\ast})]/ \partial t \rvert_{t=0}$ and $J_h\equiv\partial E [ \{A-\pi(X)\}\psi(O; \theta,h^{\ast}_{t})]/ \partial t \rvert_{t=0}$. Following  \cite{ye2021geniusmawii}, we make repeated use of the equalities (\ref{identity_1})--(\ref{identity_2}) as well as the score properties 
\begin{eqnarray}
	\label{identity_4}	&&E\{\sigma_1(X)S(A|X)\}=0;\\
	\label{identity_5}   && E\{\sigma_2(A,X)S(M|A,X)\}  = 0; \\
	\label{identity_6}	&& E\{\sigma_3(M,A,X)S(Y|M,A,X)\}  = 0, 
\end{eqnarray}
for any $\sigma_j(\cdot)$, $j=1,2,3$. The adjustment term arising from nonparametric estimation of $\pi(x)$ is
\begin{equation*}
\begin{split}
J_{\pi}&= -E[ E\{AS(A|X)|X\}\psi(O;\theta,h^{\ast})]\\
&=-E[AS(A|X)E\{\varphi(O;\theta,h^{\ast})|X\}]\\
&=-E[\{A-\pi(X)\}E\{\varphi(O;\theta,h^{\ast})|X\}S(A|X)]\\
&=-E[\{A-\pi(X)\}E\{\varphi(O;\theta,h^{\ast})|X\}S(O)]\quad (\text{\ref{identity_1}},\text{\ref{identity_5}},\text{\ref{identity_6}})\\
&=-E[\{A-\pi(X)\}\{g^{\ast}(X),\rho(X),h^{\ast}(X)\}^\T S(O)].
\end{split}
 \end{equation*}
On the other hand, the adjustment term arising from nonparametric estimation of $h^{\ast}(x)$ is
 \begin{equation*}
\begin{split}
J_{h}&=  -(0,E[\{A-\pi(X)\}(Y-\theta_1 M)E\{MS(M|A,X)|A,X\}],0)^\T\\
&=  -(0,E[\{A-\pi(X)\}E\{(Y-\theta_1 M)|A,X\}MS(M|A,X)],0)^\T\\
&=  -(0,E[\{A-\pi(X)\}\{\theta_2 A+g^{\ast}(X)\}MS(M|A,X)],0)^\T\\
&=  -(0,E[\{A-\pi(X)\}\{\theta_2 A+g^{\ast}(X)\}\{M-\theta_3 A-h^{\ast}(X)\}S(M|A,X)],0)^\T\quad (\text{\ref{identity_5}})\\
&=  -(0,E[\{A-\pi(X)\}\{\theta_2 A+g^{\ast}(X)\}\{M-\theta_3 A-h^{\ast}(X)\}S(O)],0)^\T\quad (\text{\ref{identity_2}},\text{\ref{identity_6}}).
\end{split}
 \end{equation*}
Therefore 
\begin{equation*}
\begin{split}
&\partial \theta_t/\partial t|_{t=0}=-\Delta^{-1}E\left\{\tilde{\varphi}(O;\theta,\eta)S(O)\right\},
\end{split}
 \end{equation*}
and the influence function is given by the element $-\Delta^{-1}\tilde{\varphi}(O;\theta,\eta)$  \citep{newey1994asymptotic}.

\subsection*{Proof of Lemma 3.2}
{
Assume that the regularity conditions of Theorem 1A in \cite{robins1992estimating} hold for $\tilde{\varphi}(O;\theta,{\eta})$ and $\gamma(O;\theta,\eta)$. By standard Taylor expansion,
\begin{equation*}
\begin{split}
0=&n^{-1/2}\sum_i \tilde{\varphi}(O_i;\bar{\theta},\bar{\eta}(\bar{\theta}))+\biggr[E\left\{\frac{\partial}{\partial \theta} \tilde{\varphi}(O;{\theta},\bar{\eta}(\bar{\theta}))\Big \rvert_{\theta=\bar{\theta}} \right\}-E\left\{\frac{\partial}{\partial \eta} \tilde{\varphi}(O;\bar{\theta},{\eta})\Big \rvert_{\eta=\bar{\eta}(\bar{\theta})} \right\}\\
&\times E\left\{\frac{\partial}{\partial \eta} \gamma(O;\bar{\theta},{\eta})\Big \rvert_{\eta=\bar{\eta}(\bar{\theta})} \right\}^{-1}E\left\{\frac{\partial}{\partial \theta}  \gamma(O;{\theta},\bar{\eta}(\bar{\theta}))\Big \rvert_{\theta=\bar{\theta}} \right\}\biggr]n^{1/2}(\hat{\theta}-\bar{\theta})\\
&-E\left\{\frac{\partial}{\partial \eta} \tilde{\varphi}(O;\bar{\theta},{\eta})\Big \rvert_{\eta=\bar{\eta}(\bar{\theta})} \right\}E\left\{\frac{\partial}{\partial \eta} \gamma(O;\bar{\theta},{\eta})\Big \rvert_{\eta=\bar{\eta}(\bar{\theta})} \right\}^{-1}\gamma(O_i;\bar{\theta},\bar{\eta}(\bar{\theta}))+o_p(1)\\
\equiv& n^{-1/2}\sum_i \tilde{\varphi}(O_i;\bar{\theta},\bar{\eta}(\bar{\theta}))+E\left\{\frac{\partial}{\partial \theta} \tilde{\Phi}(O;\theta,\bar{\eta}(\bar{\theta}))\Big\rvert_{\theta=\bar{\theta}}\right\}n^{1/2}(\hat{\theta}-\bar{\theta})\\
&-E\left\{\frac{\partial}{\partial \eta} \tilde{\varphi}(O;\bar{\theta},{\eta})\Big \rvert_{\eta=\bar{\eta}(\bar{\theta})} \right\}E\left\{\frac{\partial}{\partial \eta} \gamma(O;\bar{\theta},{\eta})\Big \rvert_{\eta=\bar{\eta}(\bar{\theta})} \right\}^{-1}\gamma(O_i;\bar{\theta},\bar{\eta}(\bar{\theta}))+o_p(1),
\end{split}
 \end{equation*}
 where $\bar{\theta}$ and $\bar{\eta}(\theta)$ denotes the probability limits of $\hat{\theta}$ and $\hat{\eta}(\theta)$ respectively. Assuming $$E\left\{\partial  \tilde{\Phi}(O;\theta,\bar{\eta}(\bar{\theta}))/ \partial \theta\rvert_{\theta=\bar{\theta}}\right\},$$ is non-singular, it follows that
 \begin{equation}
\begin{split}
\label{eq:Taylor}
n^{1/2}&(\hat{\theta}-\bar{\theta})=-E\left\{\frac{\partial}{\partial \theta} \tilde{\Phi}(O;\theta,\bar{\eta}(\bar{\theta}))\Big\rvert_{\theta=\bar{\theta}}\right\}^{-1} n^{-1/2}\sum_i \biggr[\tilde{\varphi}(O_i;\bar{\theta},\bar{\eta}(\bar{\theta})) \\
&-E\left\{\frac{\partial}{\partial \eta} \tilde{\varphi}(O;\bar{\theta},{\eta})\Big \rvert_{\eta=\bar{\eta}(\bar{\theta})} \right\}E\left\{\frac{\partial}{\partial \eta} \gamma(O;\bar{\theta},{\eta})\Big \rvert_{\eta=\bar{\eta}(\bar{\theta})} \right\}^{-1}\gamma(O_i;\bar{\theta},\bar{\eta}(\bar{\theta}))\biggr] +o_p(1).
\end{split}
 \end{equation}
 
 Let $\bar{\pi}(\cdot)=\pi[\cdot;\bar{\eta}_1(\bar{\theta})]$, $\bar{g}^{\ast}(\cdot)=g^{\ast}[\cdot;\bar{\eta}_2(\bar{\theta})]$, $\bar{\rho}(\cdot)=\rho[\cdot;\bar{\eta}_3(\bar{\theta})]$ and $\bar{h}^{\ast}(\cdot)=h^{\ast}[\cdot;\bar{\eta}_4(\bar{\theta})]$ denote the working models evaluated at $\bar{\eta}(\bar{\theta})$. We consider the following three cases:
\begin{itemize}
\item[(i)] If $(\bar{\theta},\bar{\pi},\bar{g}^{\ast})=(\theta^{\dag},\pi,g^{\ast})$, then 
 \begin{equation*}
\begin{split}
E\{ \tilde{\varphi}(O;\theta^{\dag},\bar{\eta}({\theta}^{\dag}))\}=& E\{\tilde{\varphi}(O;\theta^{\dag},\pi,g^{\ast},\bar{\rho},\bar{h}^{\ast})\}\\
=& E\{\tilde{\varphi}(O;\theta^{\dag},\pi,g^{\ast},\bar{\rho},\bar{h}^{\ast})\}+E[\{A-\pi(X)\}\sigma^{\ast}_1(X)]\\
&+E[\{Y-\theta^{\dag}_1 M-\theta^{\dag}_2 A-g^{\ast}(X)\}\sigma^{\ast}_2(A,X)]\quad (\text{\ref{identity_1}},\text{\ref{identity_3}})\\
=& E\{\tilde{\varphi}(O;\theta^{\dag},\pi,g^{\ast},{\rho},{h}^{\ast})\}\\
=&0,
\end{split}
 \end{equation*}
 where $\sigma^{\ast}_1(X)\equiv [0,\{\bar{\rho}^{\ast}(X)-\rho^{\ast}(X)\},\{\bar{h}^{\ast}(X)-h^{\ast}(X)\}]^\T$ and $\sigma^{\ast}_2(A,X)\equiv [0,\{A-\pi(X)\}\{\bar{h}^{\ast}(X)-h^{\ast}(X)\},0]^\T$.
\item[(ii)] If $(\bar{\theta},\bar{\pi},\bar{h}^{\ast})=(\theta^{\dag},\pi,h^{\ast})$, then 
 \begin{equation*}
\begin{split}
E\{ \tilde{\varphi}(O;\theta^{\dag},\bar{\eta}({\theta}^{\dag}))\}= & E\{\tilde{\varphi}(O;\theta^{\dag},\pi,\bar{g}^{\ast},\bar{\rho},h^{\ast})\}\\
=& E\{\tilde{\varphi}(O;\theta^{\dag},\pi,\bar{g}^{\ast},\bar{\rho},h^{\ast})\}+E[\{A-\pi(X)\}\sigma^{**}_1(X)]\\
&+E[\{M-\theta^{\dag}_3 A-{h}^{\ast}(X)\}\sigma^{**}_2(A,X)]\quad (\text{\ref{identity_1}},\text{\ref{identity_3}})\\
=& E\{\tilde{\varphi}(O;\theta^{\dag},\pi,g^{\ast},{\rho},{h}^{\ast})\}\\
=&0,
\end{split}
 \end{equation*}
 where $\sigma^{**}_1(X)\equiv [\{\bar{g}^{\ast}(X)-g^{\ast}(X)\},\{\bar{\rho}^{\ast}(X)-\rho^{\ast}(X)\},0]^\T$ and $\sigma^{**}_2(A,X)\equiv [0,\{A-\pi(X)\}\{\bar{g}^{\ast}(X)-g^{\ast}(X)\},0]^\T$.
\item[(iii)] Lastly if $(\bar{\theta},\bar{g}^{\ast},\bar{\rho},\bar{h}^{\ast})=(\theta^{\dag},g^{\ast},\rho,h^{\ast})$, then 
 \begin{equation*}
\begin{split}
E\{ \tilde{\varphi}(O;\theta^{\dag},\bar{\eta}({\theta}^{\dag}))\}= & E\{\tilde{\varphi}(O;\theta^{\dag},\bar{\pi},{g}^{\ast},{\rho},h^{\ast})\}\\
=& E[E\{\tilde{\varphi}(O;\theta^{\dag},\bar{\pi},{g}^{\ast},{\rho},h^{\ast})|A,X\}]\\
=&0.
\end{split}
 \end{equation*}
\end{itemize}

Combining the above results,  we have $E\{ \tilde{\varphi}(O;\theta^{\dag},\bar{\eta}({\theta}^{\dag}))\}=0$ if one, but not necessarily more than one, of (i), (ii) or (iii) holds, i.e. in the union model $\{\cup_{j=1}^3 \mathcal{M}_j\}$.  The asymptotic distribution of $n^{1/2}(\hat{\theta}-{\theta}^{\dag})$ follows from (\ref{eq:Taylor}) by Slutsky's Theorem and the Central Limit Theorem.

}

 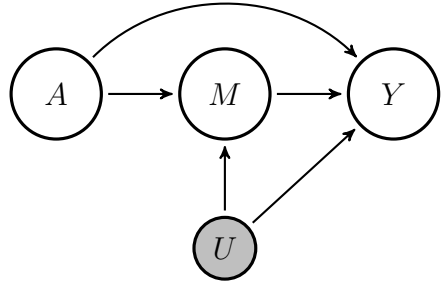
\begin{figure}[!p]
\vspace*{10pt}
		\centering	
	\begin{tikzpicture}[->,>=stealth',node distance=1cm, auto,]
	\node[est] (D) {$A$};
	\node[est, right = of D] (M) {$M$};
	\node[est, right = of M] (Y) {$Y$};
	\node[shade, below = of M] (U) {$U$};
	\path[pil] (D) edgenode {} (M);
	\path[pil] (M) edgenode {} (Y);
	\path[pil] (U) edgenode {} (M);
	\path[pil] (U) edgenode {} (Y);
	\path[pil] (D) edge [bend left=45] node {} (Y);
	\end{tikzpicture}
	\caption{Causal diagram with unmeasured mediator-outcome
confounding within strata of  $X$.}
          \label{fig1}
\end{figure}

\newpage

	\begin{table}[!p]
\centering

\def~{\hphantom{0}}
\caption{Summary of results for estimation of NDE and NIE  for a change of exposure value from $0$ to $1$ ($n=800)$.}

{%
\begin{tabular}{ccccccccccccc}
 \toprule
&\multicolumn{3}{c}{Scenario (i)} & \multicolumn{3}{c}{Scenario (ii)}&\multicolumn{3}{c}{Scenario (iii)} & \multicolumn{3}{c}{Scenario (iv)}\\
              	     \cmidrule(lr){2-4}\cmidrule(lr){5-7}\cmidrule(lr){8-10}\cmidrule(lr){11-13}
	  & \texttt{MR} & \texttt{PS} & \texttt{BK}  & 
	  \texttt{MR} & \texttt{PS} & \texttt{BK}  &
	  \texttt{MR} & \texttt{PS} & \texttt{BK}  &
	  \texttt{MR} & \texttt{PS} & \texttt{BK}  \\
	     \midrule
	     \multicolumn{13}{c}{NDE}\\
	               Bias  & .005 & .009 & $-.501$  & .006 & $-.074$ & $-.546$& .007 & .009 & $-.501$ & $.082$ & $.282$ & $-.501$ \\
$\sqrt{\text{Var}}$   & .102 & .112 & .027            & .104 & .110 & .040& .107 & .112 &.027 & $.105$ & $.162$ & $.027$ \\
$\sqrt{\text{EVar}}$ & .103 & .115 & .028           & .105 & .112 & .044& .108 & .115 &.028  & $.107$ & $.169$ & $.028$ \\
	            Cov90 & .910 & .901 & .000             & .908 & .810 & .000& .910 & .901 &.000& $.820$ & $.477$ & $.000$ \\
                       Cov95 & .949 & .944 & .000           & .952 & .875 & .000& .957 & .944 &.000  & $.904$ &  $.655$ & $.000$\\ 
          \multicolumn{13}{c}{NIE}\\
          	               Bias  & .004 & .000 & .741 & .003 & .083 & .885&  .002 & .000 &.898 &$.079$ & $-.121$ & $.741$ \\
$\sqrt{\text{Var}}$   & .167 & .174 & .191 & .168 & .178 & .209 & .168 & .174 &.204 & $.171$ & $.195$ & $.191$ \\
$\sqrt{\text{EVar}}$ & .172 & .179 & .208 & .173 & .182 & .218& .174 & .179 &.217&$.176$ & $.203$ & $.208$ \\
	            Cov90 & .918 & .912 & .019  & .915 & .876 & .005& .906 & .912 &.007 & $.873$ & $.858$ & $.019$ \\
                       Cov95 & .956 & .954 & .036  & .954 & .925 & .011&  .956 & .954 &.011& $.929$ &  $.920$ & $.036$\\ 

\bottomrule
\end{tabular}}
\label{tab1}
   \begin{tablenotes}
      \small
      \item Note: Bias and $\sqrt{\text{Var}}$ are the Monte Carlo bias and standard deviation of the point estimates, $\sqrt{\text{EVar}}$ is the square root of the mean of the variance estimates and Cov90 or Cov95 is the coverage proportion of
the 90\% or 95\% Wald confidence interval respectively, based on 1000 replicates.
    \end{tablenotes}

\end{table}

 \begin{table}[!p]
\centering

\def~{\hphantom{0}}
\caption{Estimates ($\pm$ $1.96\times$standard error) of NDE and NIE of self-efficacy on fatigue mediated though PTSD symptoms, for a change in self-efficacy from below the sample median of total scores on the General Self-Efficacy Scale $(A=0)$ to above $(A=1)$. }{
\begin{tabular}{ccccccccccccc}
 \toprule
 &\texttt{MR} &  \texttt{PS} &  \texttt{BK} \\
	     \midrule
NDE &  $-.765\pm .340$  &   $-.748\pm .332$  & $-.426\pm .146$  \\
NIE &  $\phantom{-}.066 \pm .308$   &   $\phantom{-}.049\pm .297$  & $-.273\pm .084$ \\
\bottomrule
\end{tabular}}
\label{tab2}
\end{table}
\end{document}